\documentclass[12pt]{article}
\usepackage{a4}
\usepackage{amsmath}
\usepackage{amssymb}

\begin{document}

\begin{center}

{\Large\bf Supersymmetry - Early Roots That Didn't Grow}

\vspace{.3cm}
{\bf Cecilia Jarlskog}\\

Division of Mathematical Physics, Physics Department, LTH,\\
Lund University, Lund, Sweden
\end{center} 

\vspace{.5cm}

\begin{abstract}
\noindent
This article is about early roots of supersymmetry, as found in the literature from 1940s and early 1950s. There were models
where the power of ``partners'' in alleviating divergences in quantum field theory was recognized. 
However, other currently known remarkable features of supersymmetry, such as its role in the extension of the 
Poincar\'{e} group, were not known. There were, of course, no supersymmetric non-abelian quantum field theories in those days.
This article is organized as follows:

\tableofcontents
  
\end{abstract}

\section{Introduction}

During several years I was collecting material to produce a book which was subsequently published under the title
``Portrait of Gunnar K\"{a}ll\'{e}n: A Physics Shooting Star and Poet of Early Quantum Field Theory'' \cite{GKbok}.
I found to my great surprise that K\"{a}ll\'{e}n's very first published paper \cite{gk49} 
had a fundamental element of supersymmetry in it
 - the fact that spin-$\frac{1}{2}$ fermions and spin-$0$ bosons, taken together in equal numbers of degrees of freedom,
 give a far better-behaved field theory! K\"{a}ll\'{e}n's
paper was published when he was a 23 year old second year doctoral student in Lund, Sweden. He had been sent to Pauli
in Z\"{u}rich, Switzerland, to attend Pauli's summer course of 1949. The acknowledgement of    
K\"{a}ll\'{e}n's paper reads as follows:

\begin{quote}
``I want to express my respectful gratitude to professor W. Pauli, Z\"{u}rich, who has
suggested this investigation to me, and to thank him and Dr. R. Jost and Dr. J. 
Luttinger for many helpful discussions.''
\end{quote}

\section{K\"{a}ll\'{e}n's Work}

K\"{a}ll\'{e}n considers the vacuum polarization in quantum electrodynamics (QED)
to orders $e^{2}$ and $e^{4}$, in a
model containing not only $n$ Dirac fields but also $N$ spin-$0$ particles,  all
with charge $e$ and interacting with an external electromagnetic field. He informs the readers  
that the $e^2$ case has been treated (and in fact published in the same year, 1949) by 
Jost and Rayski \cite{jost}. These authors, who were both in Z\"{u}rich, in their paper
``express their gratitude to Professor W. Pauli for his kind interest in this work as well as for
much valuable aid''.
K\"{a}ll\'{e}n informs the reader that Jost and Rayski 
have shown that to the order $e^{2}$ the
``non-gauge invariant (and divergent) terms compensate each other if one uses a suitable
mixture of spinor and scalar fields.'' The mixture being 
\begin{equation}
 N= 2n
\end{equation}
\begin{equation}
 \sum_{i=1}^{N} M_{i}^{2} = 2 \sum_{i=1}^{n} m_{i}^{2}
\end{equation}
\noindent where $M_{i}$ ($m_{i}$) denote the masses of the scalars (spinors).
The first relation, that there be as many bosonic as fermionic degrees of freedom,
is exactly the basic ingredient of supersymmetry. The second relation includes
the degeneracy condition of supersymmetry $M_{i} = m_{i}$ but it is more general. 
K\"{a}ll\'{e}n studies what happens at the order $e^{4}$, which is, of course, a much more 
difficult task. After impressive calculations, he finds that the first relation
alone is enough for removal of divergences in the fourth order. There are no divergences
in higher than the fourth order and an overall charge renormalization is all that is needed.

\vspace{.2cm}
\noindent
The paper by Jost and Rayski \cite{jost} gives credit to work published in the previous year by 
Umezawa et al. \cite{umezawa48} and by Rayski \cite{rayski48} for having shown that: 

\begin{quote}
``by coupling the electromagnetic field to several
charged fields of spinor and scalar (or pseudoscalar) types the gauge invariance of the vacuum polarization
and the vanishing of the ph.s.e. [meaning photon self-energy] may be achieved (at least in the second order of
approximation $e^{2}$) due to some compensations. It is premature to decide whether such compensations posess any
deeper meaning or whether they should be considered merely as accidental''.
\end{quote} 

\noindent
Jost and Rayski do not discuss the ``partners'' 
any further but deal with comparison of charge renormalization of spinors and scalars in this scenario. 

\vspace{.2cm}
\noindent
It should be emphasized that the self-energy of the photon was a major issue in those days.
Indeed a non-zero photon mass could easily emerge from calculations, as was shown in a seminal paper by Wentzel
\cite{wentzel}. 

\vspace{.2cm}

\noindent
To conclude this section, Pauli was very pleased with K\"{a}ll\'{e}n's work. He expressed his appreciation in several letters
(to Rudolf Peierls, Oskar Klein and others).
As mentioned before, K\"{a}ll\'{e}n's work had started when he
was attending Pauli's summer course in Z\"{u}rich. Pauli  
was a major ``pole of attraction'' to many distinguished physicists as well as young researchers.

\section{Regularization and renormalization}

An insight into the situation in 1949 is obtained by considering what was going on in  Z\"{u}rich.
Indeed the previous two years had been extremely turbulent and exciting in the world of physics, 
due to two remarkable experimental discoveries crying
for theoretical explanations. These were:

\begin{quote}
\noindent 
(1) the discovery in 1947 of the Lamb-shift \cite{lamb-shift} by Lamb\footnote{In 1955 there were two Nobel Laureates
in Physics, one of them being Willis Lamb who received the Prize for this discovery. 
In his Nobel lecture, he mentions that this effect had been seen about ten years earlier by other researchers \cite{lamb-talk} 
but unfortunately, as he puts it, had not been taken seriously. This was
perhaps fortunate for him as he could do a much more accurate measurement, using newly developed microwave techniques.

} 
and Retherford\footnote{Unfortunately sometimes in the scientific literature
one sees Retherford incorrectly spelled as Rutherford, thereby giving the credit to the latter who had passed away
about ten years earlier, in 1937. Robert Curtis Retherford (1912 - 1981) was a graduate student of Willis Lamb.

}.

\vspace{.2cm}

\noindent 
(2) the discovery in 1948 of the anomalous magnetic moment of the electron \cite{kusch} by Foley and 
Kusch\footnote{Polykarp Kusch was the second Nobel Laureate in Physics in 1955, for ``his precision determination of
the magnetic moment of the electron''.

}. Note that this anomaly was quickly calculated by Schwinger 
who announced his famous factor $\frac{\alpha}{2\pi}$ in a one-page letter \cite{schwinger48}
which appeared in Physical Review, next to the experimental discovery. 

\end{quote}

\noindent
These discoveries, which showed departure from predictions of Dirac theory, attracted the attention of some of the greatest
theorists of that time, among them H. A. Bethe, R. P. Feynman, N. M. Kroll, J. Schwinger, S-I. Tomonaga and
V. F. Weisskopf. 
Through these efforts, regularization and renormalization took the central stage in relativistic quantum field theory.

\vspace{.2cm}
\noindent
Returning to Z\"{u}rich, Pauli who was keen on writing review-type
articles, going through the details in order to understand what is going on, was quick to publish a paper
in 1949, together with F. Villars
\cite{villars}. This paper, published under the title `` On the Invariant Regularization in 
Relativistic Quantum Theory'', came to play a major role in the thinking of field theorists
for decades to come. The paper begins by noting that 

\begin{quote}
``In spite of many successes of the new relativistically invariant formalism of 
quantum electrodynamics, which is based on the idea of ``renormalization'' of mass and charge, there are
still some problems of uniqueness left, which need further clarification.''
\end{quote}   
 
\vspace{.2cm}
\noindent
The authors then point out the inherent ambiguities in calculations in quantum 
electrodynamics, due to presence of divergent integrals. Actually a number of prescriptions had been
put forward on how
to deal with these infinities, usually by introducing cut-offs\footnote{See, for example, Feynman's article  
``Relativistic Cut-Off for Quantum
Electrodynamics'' \cite{feynman} where he mentions several by then existing cut-off procedures, some
similar to his own, especially that of F. Bopp \cite{bopp}.

}. 
Regularization and renormalization were the key concepts and it was essential that the regularization prescription 
should respect Lorentz as well as gauge invariance.
 
\vspace{.2cm}
\noindent
In their article, Pauli and Villars 
specify two distinct paths to regularization:

\begin{quote}
 ``In order to overcome these ambiguities we apply in the following the method of regularization ... with the help
of an introduction of auxiliary masses. This method has already a long history. Much work has been done
to compensate the infinities in the self-energy of the electron with the help of auxiliary fields ...
Some authors assumed formally a negative energy of the free auxiliary particles, while others did not
need these artificial assumptions and could obtain the necessary compensations by using the 
different sign of the self-energy of the electron due to its interaction with different kinds of
fields (for instance scalar fields {\it vs.} vector fields).'' 
\end{quote}

\vspace{.2cm}
\noindent
Pauli and Villars call the first kind of regularization (involving fictitious particles) ``{\bf formalistic}''. Some of the 
hypothetical particles thus introduced are ghosts (with negative probabilities) 
and their masses are taken to go to infinity at the end of computation in order to remove them from
the physical sector. This formalistic prescription amounted to introducing some simple relations which
nowadays can be found in many textbooks on field theory,
$$\sum_{i} C_{i} =0; ~ \sum_{i} C_{i} M_{i}^{2}= 0, ...$$ 

\vspace{.2cm}
\noindent
Pauli and Villars call the second option ``{\bf realistic}''. In this case\footnote{Pauli and Villars, in their
paper, give credit to Rayski as an inventor of
auxiliary fictitious particles. The authors state: ``Rayski made this proposal in the summer of 1948
during his investigations on the photon self-energy of {\it Bosons} (see reference 6). With his
friendly consent we later resumed his work and generalized the method for arbitrary external
fields (not necessarily light waves)''.

\vspace{.1cm}

\noindent Note, however, that Pauli and Villars consistently refer 
to J. Rayski as G. Rayski. This could be due to the fact that he, when he had been in USA, had
translated his first name, Jerzy, into English and had published under the name George Rayski. 
Jerzy Rayski (1917-1993) from Krakow, Poland, had spent one year during the period 1949-1950
in Z\"{u}rich. He relates some of his recollections in \cite{rayski}.

}
Nature herself provides 
the regulators, in the form of what in supersymmetry language is called the partners. In comparing the two
approaches they note:

\begin{quote}
 ``It seems very likely that the ``formalistic'' standpoint used in this paper and by other workers can
only be a transitional stage of the theory, and that the auxiliary masses will eventually either be
entirely eliminated, or the ``realistic'' standpoint will be so much improved that the theory will not
contain any further accidental compensations.''
\end{quote}

\vspace{.2cm}
\noindent
The first (formalistic) option, is indeed easy to use and the only one which is used in the Pauli-Villars
paper. An immediate question
which arises is: what happened to the second alternative - the ``realistic standpoint'' - fermions accompanied by their bosonic
partners? 

\section{Yukawa, the intellectual father of scalars}

In 1940s and 1950s scalar particles enjoyed a great deal of popularity among theorists in Japan.
This can be traced back to the work of H. Yukawa\footnote{Yukawa received the 1949 Nobel Prize in Physics 
``for his prediction of the existence of mesons on the basis of theoretical work on nuclear forces''.

} 
who proposed that the strong force is mediated by a scalar particle (and its antiparticle), $U^{\pm}$.
Using Wigner's estimate of the range of the nuclear force, he found
that the mass of $U$ should be about $200$ times the mass of the electron (see Yukawa's Nobel lecture \cite{yukawaNobelLecture}).

\vspace{.2cm}
\noindent
The sequence of event from 1935 to 1949, when Yukawa was awarded the Nobel Prize, as well as what
happened afterwards, is truly amazing.
Indeed particles with the expected mass and charge were there - the muons,
$\mu^{\pm}$ (initially called the
mesotrons, later on the mu-mesons). However, these penetrating particles turned out to have no strong interactions! The
pions were discovered much later, in 1947 (for a more detailed discussion see below). 

\vspace{.2cm}
\noindent At the time of his Nobel Prize Yukawa was well aware that the picture was 
far more complicated. In his Nobel Lecture he noted \cite{yukawaNobelLecture}:

\begin{quote}
``In this way, meson theory has changed a great deal during these fifteen
years. Nevertheless, there remain still many questions unanswered. Among
other things, we know very little about mesons heavier than $\pi$-mesons. We
do not know yet whether some of the heavier mesons are responsible for
nuclear forces at very short distances. The present form of meson theory is
not free from the divergence difficulties, although recent development of
relativistic field theory has succeeded in removing some of them. We do not
yet know whether the remaining divergence difficulties are due to our 
ignorance of the structure of elementary particles themselves\cite{yukawareference20}. We shall 
probably have to go through another change of the theory, before we shall be
able to arrive at the complete understanding of the nuclear structure and of
various phenomena, which will occur in high energy regions.''
\end{quote}

\noindent
Little did Yukawa, or anyone else, know about quantum chromodynamics (QCD) and its gluons as mediators of the strong force. 
In spite of all this, light mesons play an important role in the treatment of strong interactions
at very low energies (chiral dynamics). 

\section{Scalars in Japan and ``mixed-fields''}

In 1940s there were several towering figures in the Japanese theoretical community, among them
S. Sakata (1911-1970), S-I. Tomonaga (1906-1979) and H. Yukawa (1907-1981). Following Yukawa,
proposing new scalars became popular in Japan.
As a token of importance of scalars in Japan we note that the 1965 Nobel Laureate in Physics, 
Tomonaga\footnote{The 1965 Nobel Prize in Physics was awarded to 
Sin-Itiro Tomonaga, Julian Schwinger and Richard P. Feynman ``for their fundamental work in quantum electrodynamics, 
with deep-ploughing consequences for the physics of elementary particles''.

}, in spite of not having been directly involved,
discussed this subject in his Nobel lecture \cite{tomonaganobel}. He wrote:

\begin{quote}
``In the meantime, in 1946, Sakata \cite{sakata46} proposed a promising method of eliminating the divergence of the electron 
mass by introducing the idea of a field of cohesive force. It was the idea that there exists unknown field, of the 
type of the meson field which interacts with the electron in addition to the electromagnetic field. Sakata named this 
field the cohesive force field, because the apparent electronic mass due to the interaction of this field and the electron, 
though infinite, is negative and therefore the existence of this field could stabilize the electron in some sense. 
Sakata pointed out the possibility that the electromagnetic mass and the negative new mass cancel each other and that 
the infinity could be eliminated by suitably choosing the coupling constant between this field and the electron. 
Thus the difficulty which had troubled people for a long time seemed to disappear insofar as the mass was concerned. 
(It was found later that Pais \cite{pais46} proposed the same idea 
in the U.S. independently of Sakata.) 

That is to say, according to our result, the Sakata theory led to the cancellation of infinities for the mass but not 
for the scattering process. It was also known that the infinity of vacuum polarization type was not 
cancelled by the introduction of the cohesive force field.''

\end{quote}
  
\noindent
Pais had an earlier publication in Physical Review \cite{pais45} in which he promised a forthcoming article which appeared
in Trans. Roy. Acad. Sciences of the Netherlands 19, no. 1 (1947).

\vspace{.2cm}
\noindent 
Returning to Japan,  
a key player in proposing ``mixed fields'' - harmonious existence of fermions and bosons -  
was H. Umezawa\footnote{For information about Umezawa see, for example, G. Vitiello on `` Hiroomi Umezawa and Quantum Field
Theory'' \cite{vitiello2011}. I wish to thank Professor Vitiello for sending his article to me.

} 
(1924-1997) and an excellent account of the status 
of the mixed fields, as of 1950, can be found in an article ``On the Applicability of the Method of
the Mixed Fields in the Theory of the Elementary Particles'' that he wrote together with Sakata \cite{umezawa50}.
Fortunately, Umezawa produced an excellent book (translated from Japanese) on ``Quantum Field Theory'' \cite{umezawaBook}
which not only teaches field theory but is also a rich source of historical information. 
Umezawa considers the case of mixed-fields in a few places in his book. As an example, the equations presented on page 255 
must look familiar to researchers in the field of supersymmetry\footnote{These being 
$$ k-2l +3m =0$$
$$ \sum_{i}^{k} (\kappa_{i}^{(s)})^{2} - 2 \sum_{i}^{l} (\kappa_{i}^{(f)})^{2} + 3 \sum_{i}^{m} (\kappa_{i}^{(v)})^{2} =0$$

\noindent
where $k, l$ and $m$ denote the number of (charged) scalar, spin-$\frac{1}{2}$, and spin-1 fields respectively and
$\kappa_{i}$ denotes the appropriate mass. 

}! Taken altogether, one is impressed by 
the amount of work done on this subject in Japan. 

\vspace{.2cm}
\noindent
The mixed fields had in turn their roots. Perhaps the earliest one was 
the introduction of a ``subtractive'' vector field,
proposed by F. Bopp \cite{bopp}. This amounted to replacing the $\frac{1}{r}$ Coulomb potential with
$\frac{1}{r} (1- exp (-\kappa r))$, or later on, in the relativistic version, introducing a cutoff in the propagator, viz, 
$\frac{1}{k^{2}} - \dfrac{1}{k^{2} - \kappa^{2}}$, $\kappa$ being the cutoff mass.

\vspace{.2cm}
\noindent
The mixed-fields cancellations worked in the case of vacuum polarization 
because all charged particles are running in the loop. So the divergence created by the singularities in the 
electron propagators is compensated by those of its scalar partners. However, in the mixed fields models
the photon did not have any fermionic partner!

\section{Interplay theory - experiment}

Carl D. Anderson (the discoverer of the positron\footnote{The Sixth General Assembly of IUPAP (the International Union of Pure 
and Applied Physics) which took place in Amsterdam, in 1948, 
unanimously recommended the use of the term electron for both $e^{\pm}$ and the terms positon and negaton for denoting
$e^{+}$ and $e^{-}$ respectively. In the long run, however, the physics community did not follow this recommendation.

}
and one of the discoverers of the muon)
has described the circumstances concerning the discovery of the positron \cite{anderson61} as follows:

\begin{quote}
`` ... it has often been stated in the literature that the discovery of the positron was a consequence of its
theoretical prediction by Dirac, but this is not true. The discovery of positron was wholly accidental. 

... The aim of the experiment that led to the discovery of the positron was simply to measure directly
the energy spectrum of the secondary electrons produced in the atmosphere and other materials by the 
incoming cosmic radiation which at that time (1930) was thought to consist primarily of a beam of photons
or gamma rays ... ''
\end{quote}

\noindent
Anderson emphasizes the importance of being familiar with theory by noting that
if one had known the Dirac theory one

\begin{quote}
 
``could have discovered the positron in a single
afternoon. The reason for this is that the Dirac theory could have provided an excellent guide as to just how
to proceed to form positron-electron pairs out of a beam of gamma-ray photons. History did not proceed in such a
direct and efficient manner, probably because the Dirac theory, in spite of its successes,
carried with it so many novel and seemingly
unphysical ideas, such as negative mass, negative energy, infinite charge density, etc. Its highly 
esoteric character was apparently not in tune with most of the scientific thinking of that day. Furthermore, 
positive electrons apparently were not needed to explain any other observations.'' 
\end{quote}

\noindent The discovery of the muon involved a long process and it took almost two decades to know its nature. Already in
1929 the so called ``penetrating radiation'' from outer space was discovered by Walther Bothe\footnote{In
1954 Walther Bothe received a Nobel Prize in Physics for ``the coincidence method and his discoveries made therewith''.

}
and W. Kolh\"{o}rster. It took a few years before one could show that the penetrating radiation consisted of new kind of
particles which were neither protons nor electrons \cite{anderson61}. Stated briefly, 
by 1939, in spite of observation of range, curvature, ionization, and penetrating power of the mesotrons 
one was not quite sure what they were \cite{anderson39}. It was found, however, that the mesotrons didn't interact much with matter
\cite{wilson39}. By 1947 it was definitively established
\cite{conversi} that the observed mesotrons were not relevant for transmitting strong interactions - a large fraction of 
negative mesotrons, instead of being captured by nuclei, decayed - the capture rates were by about 12  orders of 
magnitude smaller than expected! 
On the interplay of theory and experiment in this case Anderson says:

\begin{quote}
``This novel suggestion of Yukawa was unknown to the workers engaged in the experiments on the muon until after
the muon's existence was established. ... It is interesting to speculate on just how much Yukawa's suggestion,
had it been known, would have influenced the progress of the experimental work on muon. My own opinion is that 
this influence would have been considerable even though Dirac's theory, which was much more specific than
Yukawa's, did not have an effect on the positron's discovery. My reason for believing this is that for a period of 
almost two years there was strong and accumulating evidence for the muon's existence and it was only the caution of
the experimental workers that prevented an earlier announcement of its existence. I believe that a theoretical idea
like Yukawa's would have appealed to the people carrying out the experiments and would have provided them with a 
belief that maybe after all there is some need for a particle as strange as a muon, especially if it could help
explain something as interesting as the enigmatic nuclear force. ...'' 
\end{quote}
 
\noindent
The discovery of charged pions was announced in 1947 (and that of neutral pions a couple of years later) by researchers
working at Bristol \cite{powell} and
Cecil F. Powell received the 1950 Nobel Prize in Physics for ``his development of the photographic method of 
studying nuclear processes and his discoveries regarding mesons made with this method''.

\vspace{.2cm}
\noindent
In his Nobel Lecture \cite{powellNobelLecture}, Powell has nothing to say about any theoretical influence. He ends his talk
by stating:

\begin{quote}
 
``In the years which have passed, the study of what might, in the early days,
have been regarded as a trivial phenomenon has, in fact, led us to the discovery of many new forms on
matter and many new processes of fundamental physical importance. It has contributed to the development of a 
picture of the material universe as a system in a state of perpetual change and
flux; a picture which stands in great contrast with that of our predecessors
with their fixed and eternal atoms. At the present time a number of widely
divergent hypotheses, none of which is generally accepted, have been advanced to account for the 
origin of the cosmic radiation. It will indeed be of
great interest if the contemporary studies of the primary radiation lead us -
as the Thomsons suggested, and as present tendencies seem to indicate - to
the study of some of the most fundamental problems in the evolution of the
cosmos.''

\end{quote}

\noindent Professor Donald Perkins, in Oxford, UK, who was a member of the Bristol 
group\footnote{I wish to thank Professors Don Perkins and G\"{o}sta Ekspong, who both were among the
young members of the Bristol groups, for inspiring correspondence on
their work at Bristol. Powell refers to their work in his Nobel Lecture \cite{powellNobelLecture}, however, on his 
list of references, Ekspong's former last name (Carlson) is misspelled (as Carison).

} has confirmed
that the discovery of pions had nothing to do with theory but could simply be described by \cite{perkins}
 ``ignorant experimentalists looking for anything of
interest in emulsions exposed to cosmic radiation''. The kaons were also discovered in 1947 and that eventually led
to a new exciting period in the history of particle physics ($\theta - \tau$-puzzle, parity violation) during
which there were fruitful close contacts between theory and experiment. 

\vspace{.2cm}
\noindent
In the case of supersymmetry, there have been close contacts between theorists and experimentalists. It remains to be 
seen what the outcome of these joint efforts is going to be.
 
\section{The rich heritage of relativistic quantum field theory and its future}

Richard Feynman, in a talk given in 1979, made the following statement:

\begin{quote}
``We have a theory which is called quantum electrodynamics which is our pride and joy.'' 
\end{quote}

\noindent
Indeed, Relativistic quantum electrodynamics (QED) is one of the crown jewels of human achievements in physics. 
It emerged at the end of 1920s through efforts to describe the interactions of light quanta with electrons.
However, two decades of work by a large number of researchers was required to reformulate it and to be able to make sense
out of it. 
Naturally, there were several earlier milestones on the road that led to QED. It is perhaps appropriate to briefly remind
ourselves of a few of them: 

\vspace{.2cm}
 
\textbullet ~ Introduction of the quantum of action $h$ by Max Planck (1900)

\vspace{.2cm}

\textbullet ~ Einstein's theory of special relativity (1905)

\vspace{.2cm}

\textbullet ~ Einstein's introduction of light as a ``quantum {\bf particle}'', $E= h \nu$ (1905)

\vspace{.2cm}

\textbullet ~ ``Creation of quantum mechanics''\footnote{Quoting the Prize motivation that 
the Nobel Prize in Physics 1932 was awarded to 
Werner Heisenberg ``for the creation of quantum mechanics, ...''

} 
by Heisenberg (1925) and Schr\"{o}dinger's version of it (1926)

\vspace{.2cm}

\textbullet ~ Dirac's introduction of annihilation and creation operators for photons (1927) and his first order relativistic 
description of the electron (1928)

\vspace{.2cm} 

\textbullet ~ Quantization rules for bosons, using commutators, and for fermions, using anticommutators as well as a formal
derivation of the Pauli principle (1928)
  
\vspace{.2cm}

\noindent 
These theoretical breakthroughs took us to the end of 1920s when it was realized that  QED, in spite of all its successes, 
faced insurmountable-looking difficulties, such as infinite corrections to the energy levels of atoms.
The conclusion drawn by some researchers was that QED, of those days, will not be applicable to any problem
where relativistic effects are important.

\vspace{.2cm}
\noindent
On the status of field theory later on, one of the leading field theorist of our time, Steven Weinberg writes 
(\cite{weinberg}, p. 281): 

\begin{quote}
`` During 1930s it was widely believed that these infinities signified the breakdown of quantum electrodynamics
at energies above a few MeV. ... [after the war] ... The great success of calculations in quantum electrodynamics 
using the renormalization idea generated a new enthusiasm for quantum electrodynamics.''
\end{quote}

\vspace{.2cm}
\noindent
However by the end of 1950s and in 1960s quantum theory came under fierce attack, at least in some
theoretical circles. By that time, the number of hadrons had increased so much that it raised the question whether they
were all ``elementary particles'', each with its own field. The pion-nucleon coupling constant was found to be so large
that it did not allow physicists to apply their favorite approach - perturbation theory.
What was there to be done? Indeed, a group of theorists advocated departure from field theory altogether. It was argued that 

\begin{quote}
``there is no reason why some particles should be on a different footing from others. 
The elementary particle concept is unnecessary, at least for baryons and mesons.'' 
\end{quote}

\noindent
Some distinguished physicists declared field theory as dead and buried, as far as
hadrons were concerned. Why it worked well for quantum electrodynamics was considered as a puzzle. 
As we know from later history,  
field theory retaliated by striking back with stronger force than ever before and gave us the electroweak theory as well as QCD!  
  
\vspace{.2cm}
\noindent
What we wish to know now is whether supersymmetry is realized in nature. Viki Weisskopf liked to 
say ``predicting is difficult, especially when it concerns the future'',
a statement that he attributed to Niels Bohr. Indeed we all have ``feelings'' and ``wishes''. We would like to tell
Nature how She should behave, but She has already made up Her mind and it is for us to find out
what Her feelings and wishes have been. In any case, discovery of supersymmetry  
need not signify any departure from our current framework: relativistic local quantum field theory.

\vspace{.2cm}
\noindent
Concerning the future, I would like to quote Steven Weinberg who writes (\cite{weinberg}, p. 289):

\begin{quote}
 ``My own view is that all of the successful field theories of which we are so proud - electrodynamics, the
electroweak theory, quantum chromodynamics and even General Relativity - are in truth effective field theories,
only with a much larger characteristic energy, something like the Planck energy, $10^{19}$ GeV.''
\end{quote}

\vspace{.2cm}
\noindent
Let's hope that Mother Nature will soon reveal some of her instructive secrets, for us to be able 
to proceed further on the ``right path''.

\end{document}